\input epsf
\documentstyle{mn}
\begin{document}
\title[The distances to the X-ray binaries LSI +61$^\circ$ 303 and A0535+262]
{The distances to the X-ray binaries LSI +61$^\circ$ 303 and A0535+262}
\author[I.A. Steele et al.]
{I.A.Steele,$^1$ 
I. Negueruela,$^1$
M.J.Coe,$^2$ and
P. Roche$^3$\\
$^1$ Astrophysics Research Institute, Liverpool John Moores University, 
Liverpool, L3 3AF\\
$^2$ Department of Physics and Astronomy, 
University of Southampton, Southampton, SO17 1BJ
\\
$^3$ Astronomy Centre, CPES, 
University of Sussex, Brighton, BN1 9QJ\\
}
\maketitle
\begin{abstract}
Chevalier \& Ilovaisky (1998) use {\it Hipparcos} data to show that the
X-ray binary systems LSI+61$^\circ$ 303 and A0535+262 
are a factor of ten closer (i.e. $d\sim$ few hundred pc) than
previously thought ($d\sim2$kpc).  We present high quality
CCD spectra of the systems, and conclude that the spectral
types, reddening and absolute magnitudes of these objects are 
strongly inconsistent with the closer distances.  We propose that
the {\it Hipparcos} distances to these two systems are incorrect due to
their relatively faint optical magnitudes.
\end{abstract}

\section{Introduction}

In a recent paper
Chevalier \& Ilovaisky (1998 - CI98) presented {\it Hipparcos}
distances to 17 massive X-ray binary systems.  In particular
they presented results that appeared to indicate that two systems
(LSI +61$^\circ$ 303, A0535+262) were up to a
factor 10 closer than previously thought.  This has profound
implications for any models one constructs for these systems,
for instance suggesting they need only contain white dwarfs rather
than neutron stars to explain their X-ray luminosity.

In this paper we use CCD spectra to redetermine the
spectral type of LSI +61$^\circ$ 303 and A0535+262.
In both cases we show that the derived spectral types
and reddenings are strongly consistent with normal
Be stars at the distances previously ascribed to the systems, and not with
the new, closer distances.  Finally we discuss how the discrepancy 
between the {\it Hipparcos} and our distances may be explained in terms of
the faint nature of these particular sources.

\section{Spectral Types}

\subsection{LSI +61$^\circ$ 303}
An optical spectral type of B1Ib was assigned to LSI +61$^\circ$ 303
by Gregory et al. (1979) using low dispersion image tube spectrograph
observations.  However using an extinction
$A_V\sim3.3$ they argued that the true luminosity is probably
less than that implied by the supergiant classification.  Based
on UV line strengths Howarth (1983) also assigned a
spectral type of around B1.  However his analysis of
the extinction in this object
led him to fit a dereddened model atmosphere of temperature
$\sim15000$K, corresponding to a spectral type of B4.5III at
a distance of 2.4 kpc. 

\def\epsfsize#1#2{1#1}
\begin{figure*}
\begin{minipage}{6in}
\setlength{\unitlength}{1.0in}
\centering
\begin{picture}(6.0,3.6)(0,0)
\put(-1.0,-6.4){\epsfbox[0 0 2 2]{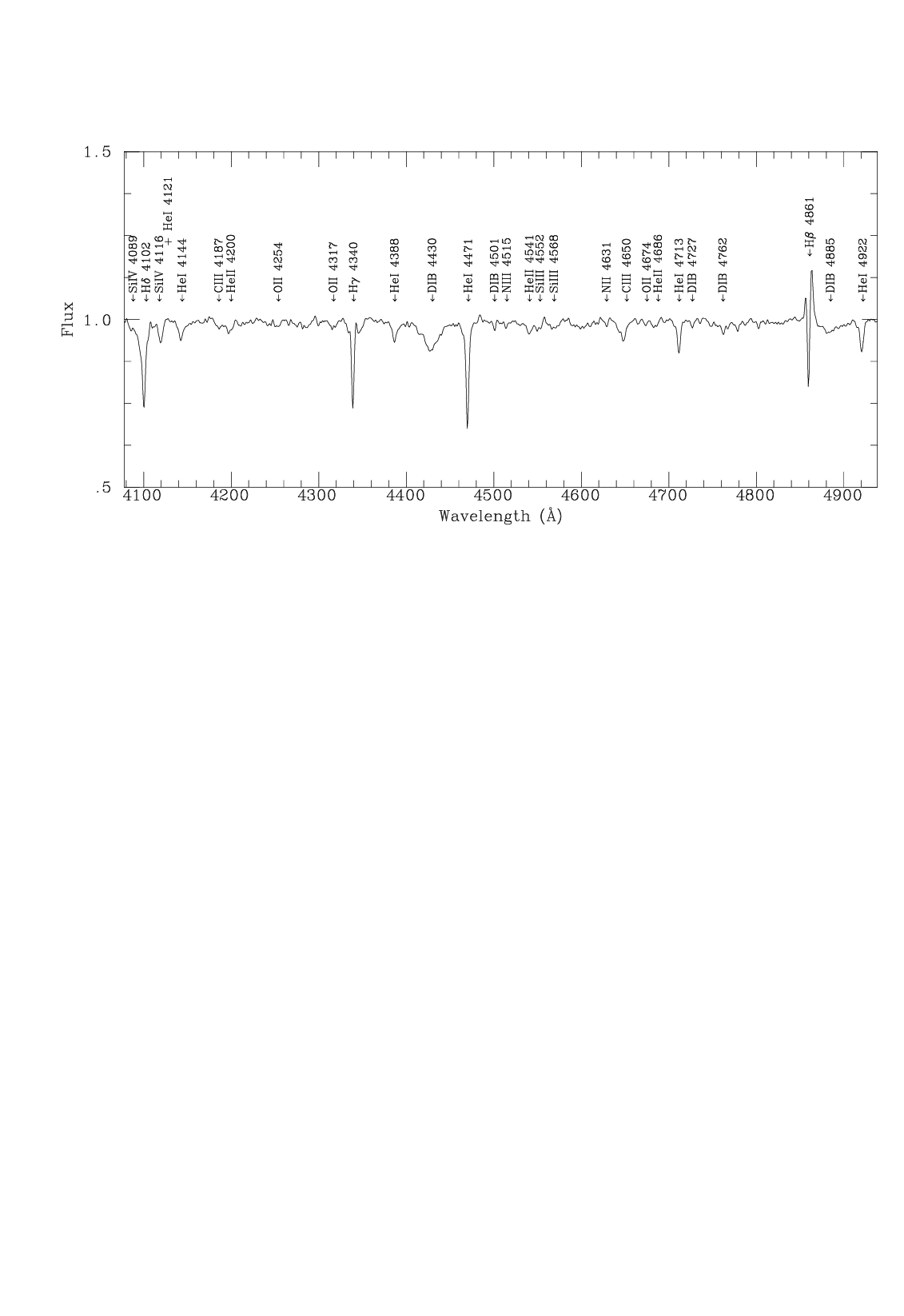}}
\end{picture}
\caption{Blue spectrum of LSI +61$^\circ$ 303.  The continuum has
been normalized to 1.0 and
a small degree of Gaussian smoothing ($\sigma=0.6${\AA}) applied.}
\end{minipage}
\end{figure*}

Fig. 1 shows a CCD spectrum of LSI+61$^\circ$303 obtained using the IDS
spectrograph of the Isaac Newton Telescope (INT), La Palma on the night of
1995 July 7 using the 1200 line/mm grating and the 235mm camera.
H$\beta$ presents a typical shell profile, but the 
line wings do not seem to be as extended as those of H$\alpha$. H$\gamma$ 
shows
a weaker shell profile, while the asymmetric shape of H$\delta$ and the
He {\sc i} lines is an
indication of an emission component on top of the underlying photospheric
feature.

The presence of the He\,{\sc ii} lines at
$\lambda\lambda$ 4200, 4541 \& 4686 \AA\ implies an early spectral type,
while their weakness indicates that
it cannot be much earlier than B0.  The Si\,{\sc iv} $\lambda\lambda$
4089 \& 4116 \AA\ lines (on the wings of H$\delta$) are very weak in 
comparison
to He\,{\sc i} $\lambda$ 4121 \AA, indicating a main-sequence object 
(Walborn \& Fitzpatrick 1990). 
A few O\,{\sc ii} lines have been tentatively identified in Fig. 1,
although none is certain.  Their absence would make the object 
earlier than B0.5. 
The lines of the Si\,{\sc iii} 
triplet are still visible, though very weak. 
The presence of C\,{\sc iii} and N\,{\sc 
iii}
lines also argues for an early type. Overall we believe that the most likely
classification is B0V, though an slightly earlier spectral type (up to O9.5V)
is also possible.

\def\epsfsize#1#2{1#1}
\begin{figure*}
\begin{minipage}{6in}
\setlength{\unitlength}{1.0in}
\centering
\begin{picture}(6.0,4.6)(0,0)
\put(-1.0,-5.5){\epsfbox[0 0 2 2]{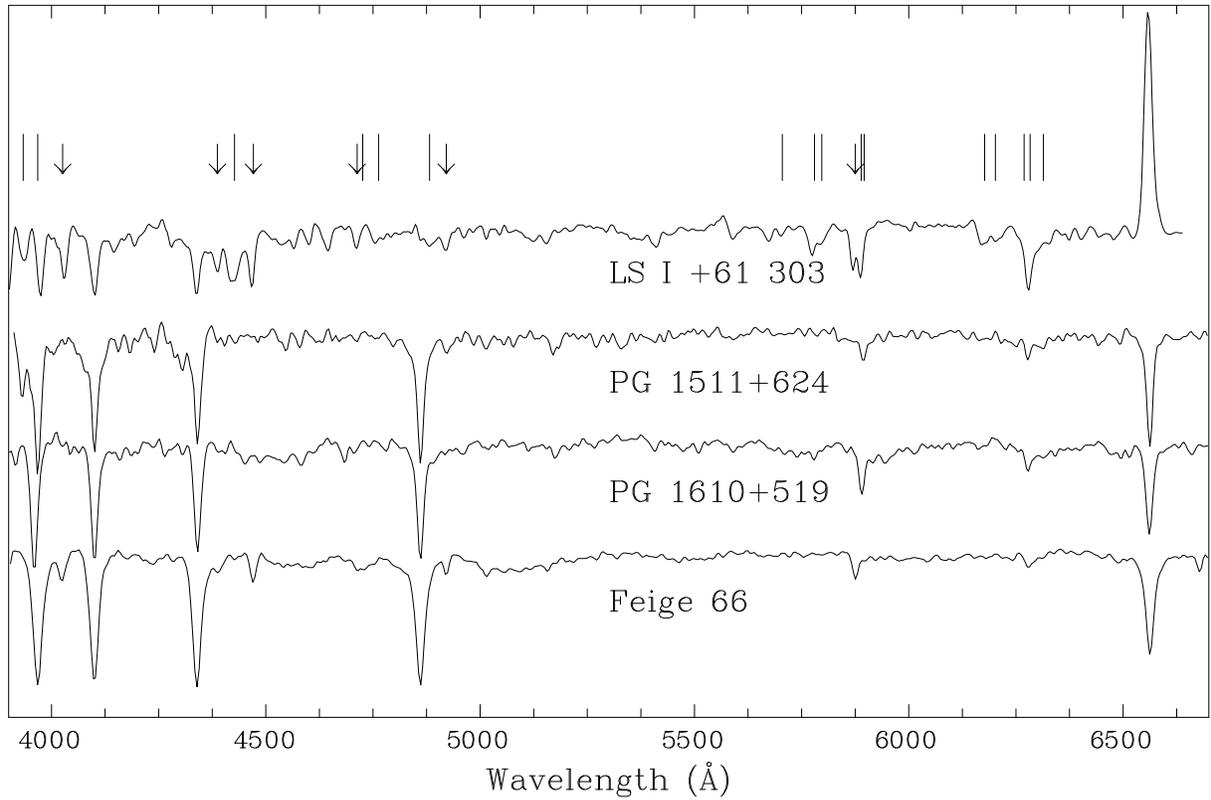}}
\end{picture}
\caption{A comparison of the spectrum of LSI+61$^\circ$ 303 and 3 
hot subdwarf (sdOB) spectra 
from the WHT data archive. The spectra have all been normalized to 1.0 and
are plotted with a constant offset.
Arrows indicate He\,{\sc I} features.  
Lines indicate the strongest interstellar features.} 
\end{minipage}
\end{figure*}

We note here that one possibility to explain a low-brightness blue star 
at a distance $\sim 200$ pc is 
that the object is a hot subdwarf. At least one X-ray binary containing a 
hot 
subdwarf and a compact object is known (HD 49798 = 1WGA J0648.0-4418; see 
Bisscheroux et al. 1997). However, it is clear that the
spectrum of LSI+61$^\circ$303 is not that of a subdwarf. The presence 
of 
helium lines prevents it from being an sdB object, while the ratio 
between 
He\,{\sc i}/He\,{\sc ii} argues against an sdO. All sdO stars show 
He\,{\sc ii}
$\lambda$4686 \AA\ $\geq$ He\,{\sc i} $\lambda$4713 \AA\ (generally, 
He\,{\sc ii}
$\lambda$4686 \AA\ $\gg$ He\,{\sc i} $\lambda$4713 \AA, see Hunger et al. 
1981)
while the opposite is true for LSI+61$^\circ$303. 

The only possibility therefore is an sdOB star.
Figure 2
compares LSI+61$^\circ$303 with three sdOB objects. 
Feige 66 is a very well-studied sdOB subdwarf (Baschek et al. 1982), 
while PG 1511+624 and PG 1610+519 have been classified as sdOB by
Moehler et al. (1990) and analysed by Allard et al. (1994), who derive 
absolute
magnitudes of $M_{V} = 6.1$ and 5.5 respectively. These magnitudes are 
compatible with their average value $\left< M_{V} = 5.9\pm 0.5 \right>$ for
hydrogen-rich subdwarfs, but are much fainter than the value we would derive
for LSI+61$^\circ$303 if it was placed at $\sim 200$pc.

There are two striking differences between the spectrum of 
LSI+61$^\circ$303
and those of the subdwarfs. First, the strong interstellar absorption 
lines, 
which are missing in the spectra of the subdwarfs, all of which are 
located at
high galactic latitudes (the implications of the strong absorption lines in
the spectrum of LSI+61$^\circ$303 are discussed in Section 4.1). 
Second, the He\,{\sc i} lines are much stronger in LSI+61$^\circ$303
even though there must be emission contamination in some lines 
(e.g. He{\sc I} 6678 {\AA} is known to show shell emission -- 
Paredes et al. 1994).  
He {\sc i} 4026{\AA}
and 4388 {\AA} are much stronger in LSI +61$^\circ$ 303 than in Feige 66
and the 5873 {\AA} He{\sc I} feature is not visable in any of the 
sdOB spectrum.
It is therefore evident that LSI+61$^\circ$303 is
a hydrogen burning star and not a post-AGB object.

\subsection{A0535+262}

\def\epsfsize#1#2{1#1}
\begin{figure*}
\begin{minipage}{6in}
\setlength{\unitlength}{1.0in}
\centering
\begin{picture}(6.0,6.6)(0,0)
\put(-1.0,-3.3){\epsfbox[0 0 2 2]{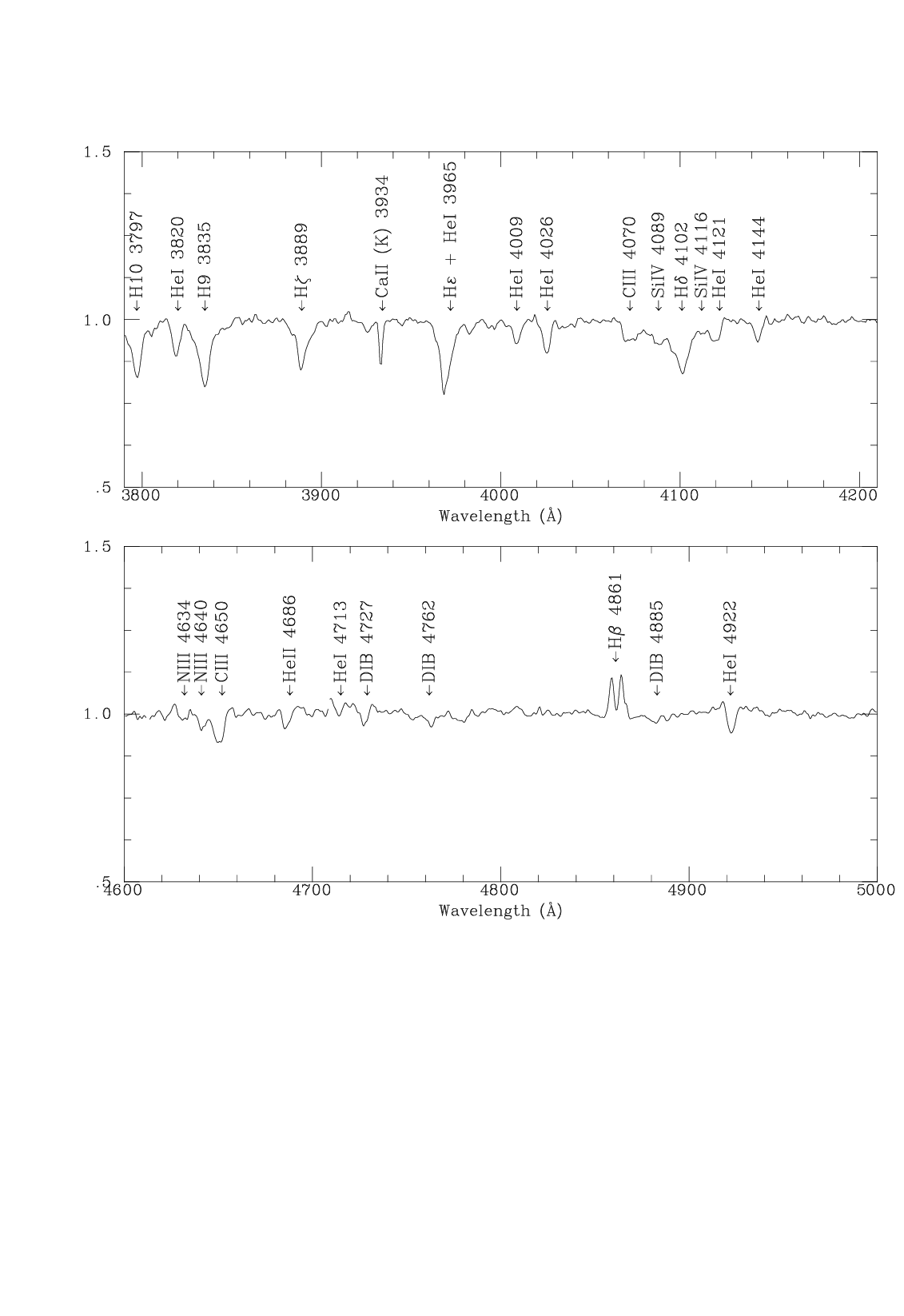}}
\end{picture}
\caption{Spectra of A0535+262 (HDE 245770).  The continuum has been
normalized to 1.0 and  
a small degree of Gaussian smoothing ($\sigma=0.6$\AA) applied.} 
\end{minipage}
\end{figure*}

The usually quoted spectral type of HDE 245770 (the optical
counterpart of A0535+262) is O9.7IIIe, based on photographic spectra
(Giangrande et al. 1980).
Fig. 3 shows two spectra of HDE 245770 
in the 
classification region.  The spectrum covering the wavelength range 
$\lambda \lambda$ 4600\,--\,5000 \AA\ was 
taken on 1996 March 1, using the Jacobus Kapteyn Telescope (JKT) 
telescope, equipped with the Richardson Brealey Spectrograph (RBS) 
(Edwin 1988) and the R2400 grating. 
The spectrum 
covering
$\lambda \lambda$ 3800\,--\,4200 \AA\ was taken on 1990 November 14,
using the IDS and the 235-mm camera on the INT, equipped with the 
R1200Y grating.

H$\beta$ presents typical double-peaked emission (see Clark et al. 1998), but
the higher Balmer lines seem to be relatively emission-free.
The presence of the He\,{\sc ii} $\lambda$ 4686 \AA\ line while He\,{\sc 
ii} 
$\lambda$ 4200 \AA\ is absent indicates a spectral type close to B0. 
Once
again, the main luminosity
criteria are the strengths of the Si\,{\sc iv} lines. Strong Si\,{\sc iv} 
$\lambda\lambda$ 4089 \& 4116 \AA\ lines are clearly visible on the wings 
of 
H$\delta$. The ratio Si\,{\sc iv} $\lambda$4089 $\sim$ He\,{\sc i} 
$\lambda$4121 is indicative of a giant star (Walborn \& Fitzpatrick 1990).
The strength of  He\,{\sc i} $\lambda$4009 \AA\ argues against a spectral
type earlier than O9.5, while the absence of O\,{\sc ii} and the presence
of N\,{\sc iii} make the object earlier than B0.5. Since both He\,{\sc ii} 
$\lambda$ 4541 \AA\ and the Si\,{\sc iii} 
triplet can be seen in lower-resolution
spectra (e.g., Clark et al. 1998) we believe that B0IIIe is the most 
appropriate classification, though the previously accepted O9.7IIIe 
cannot be discarded.

\section{Reddening}

\subsection{LSI +61$^\circ$ 303}

Howarth (1983) made an analysis of eight short wavelength and 
five long wavelength 
IUE spectra of LSI +61$^\circ$ 303.  By
flattening the {2200\AA} extinction bump he derives $E(B-V)=0.75\pm0.1$.
Such a large reddening is inconsistent with a distance of
only $\sim 200$ pc (Ishida 1969).  It is therefore important to
determine the true value of the reddening to the system.  We do this
using two methods -- 
the sodium D$_2$ line in our spectra and the strength of optical
diffuse interstellar bands.

\subsubsection{Derivation of Reddening from the Interstellar Sodium D$_2$ Line}

The first method we employ uses
the correlation between the strength of the interstellar 
Sodium D$_2$ {5890\AA} line and
extinction derived by Hobbs (1974).  We have rederived $E(B-V)$ values
for all of Hobbs (1974) objects for which he quotes Na D$_2$ equivalent
widths using spectral classifications and $B-V$ colours from
the Bright Star Catalogue (Hoffleit \& Jaschek 1982) and the intrinsic
colours of Popper (1980) for dwarfs and giants and Johnson (1966) 
for supergiants.  The correlation in plotted in Fig.4.
The scatter in this diagram is large, and
shows that this technique will not be particularly accurate in
deriving $E(B-V)$.  The dataset does not
extend beyond $E(B-V)\sim0.7$, and is sparse beyond $E(B-V)\sim0.3$.

Recognising the limitations of Fig. 4 however, we may still attempt to
use the Na D$_2$ line to determine the extinction to the system.
Two spectra of LSI +61$^\circ$ 303 were obtained in the region covering the Na
D lines on the nights of 1994 March 26 and 27 using the RBS of the 
JKT.
The mean Na D$_2$ EW measured was $650\pm90$ m\AA.  
From Figure
4 this implies $E(B-V)=0.7\pm0.4$, where the error is derived from 
the apparent spread in $E(B-V)$ in the figure.  This supports
the extinction determination of Howarth (1983).

\subsubsection
{Derivation of Reddening from Diffuse Interstellar Band Strengths}

A similar method may be employed using the diffuse interstellar bands (DIBs)
to measure extinction.  Herbig (1975) provides plots of
$E(B-V)$ versus EW for a number of diffuse bands.  These plots show
less intrinsic scatter than the Na relations of Hobbs (1974) and
should therefore give a better measure of $E(B-V)$.  In addition they
extend to higher $E(B-V)$ values ($\sim2.0$) than the Hobbs (1974)
sodium data, and so better
cover the range of interest here. 
We measure the lines in two CCD spectra taken
during the previously described JKT observing run, and a total of eight 
spectra obtained
on 1993 December 5 and 7 from the 1.5m telescope at Mount Palomar using the
f/8.75 Cassegrain echelle spectrograph in regular grating mode
(McCarthy 1988).  The DIBs
measured were those centred at 5780, 5797, 6269 and 6613 {\AA} 
(we do not employ the strong 4430 {\AA} DIB as Herbig shows that
it is only poorly correlated with $E(B-V)$).
The mean equivalent widths in each band are 480, 160, 110 and 250
m{\AA} respectively, corresponding to $E(B-V)$ values of
0.8, 0.4, 0.5 and 0.9.
The mean $E(B-V)$ derived from this method 
is therefore $0.65\pm0.25$, where the error reflects the scatter in the
values derived for the various bands.  This is again similar to 
the Howarth (1983) value.

\def\epsfsize#1#2{0.6#1}
\begin{figure}
\setlength{\unitlength}{1.0in}
\centering
\begin{picture}(3.0,2.6)(0,0)
\put(-0.3,0){\epsfbox[0 0 2 2]{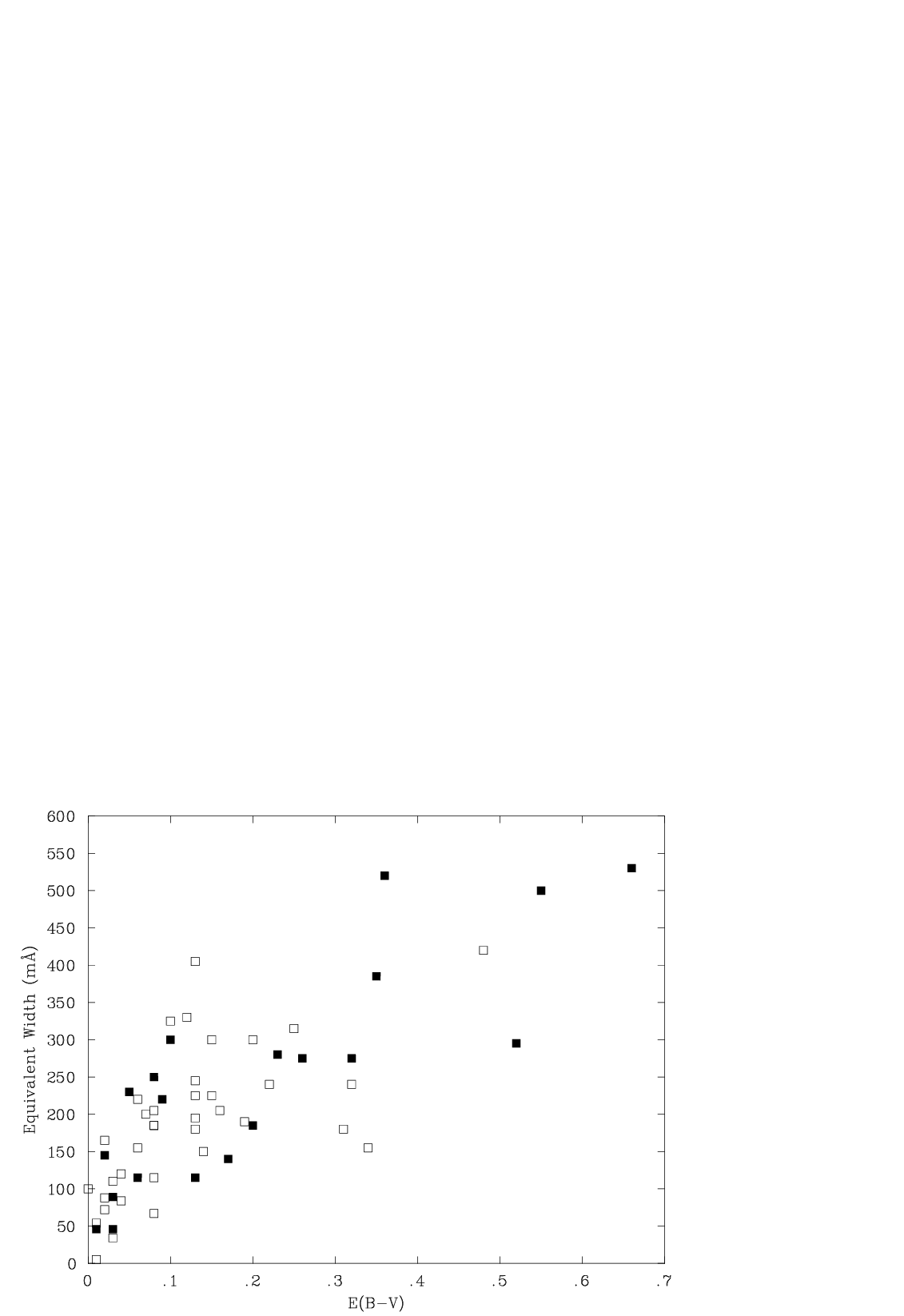}}
\end{picture}
\caption{Equivalent width of Na D$_2$ {5890\AA} versus $E(B-V)$ from the
data of Hobbs (1974).  Filled symbols indicate emission line objects,
empty symbols non-emission line objects.}
\end{figure}

\subsection{A0535+262}

The reddening to A0535+262 was determined from the 2200 {\AA} feature 
to be $E(B-V)\sim0.75$ (Giovannelli et al. 1980). Again this is larger than
one would expect for an object at $<1$kpc, and we therefore 
apply a similar analysis to the Na D$_2$ and DIB features
in spectra of A0535+262 as we did for LSI+61$^\circ$ 303.  
Using two JKT spectra obtained on 1994 March 26 and 27
we find a mean Na D$_2$ equivalent width of 500$\pm100$m{\AA}, corresponding
to an $E(B-V)=0.6\pm 0.3$.  Similarly using eight Mount Palomar
spectra from 1993 December 5 plus the two JKT spectra
we find mean DIB strengths for the 5780, 5797, 6269 and 6613 {\AA} bands
of 530, 270, 100 and 180 m{\AA} respectively, corresponding to
$E(B-V)$ values of 0.8, 0.8, 0.4 and 0.6, giving a mean $E(B-V)=0.65\pm0.2$.
Once again the similarity between all three measures (2200 {\AA} bump,
DIB, and Na D$_2$) is striking.

\section{Distances to the systems}

\subsection{LSI +61$^\circ$ 303}

The previous best estimate of the distance to LSI+61$^\circ$ 303
was that of Frail \& Hjellming (1991).  They used the profile
of the 21cm hydrogen and C$^{18}$O lines along with a galactic structure model
to place the object at a distance of 2kpc.  Using a reddening 
$E(B-V)\sim0.7$ (the average of the IUE, Na D$_2$ and DIB values) 
and the observed {\it Hipparcos} magnitude of the source, a distance of 2kpc
corresponds to an $M_V\sim-4$.  Vacca et al. (1996) quote $M_V\sim-4.2$ for
a B0V star and $M_V\sim-5.4$ for a B0III star.  Such a 
distance is therefore entirely consistent with our derived spectral
type. 

CI98 quote a {\it Hipparcos}
distance of 177 pc to the system, with 1$\sigma$ limits set at 130 and 300pc.
Assuming $E(B-V)\sim0.7$ a distance of 177pc corresponds
to an absolute $M_V \sim +2$.  The intrinsic colours derived using
such a reddening are $(B-V)_o \sim 0.05$ and $(U-B)_o \sim -0.8$.
CI98 state that these colours are not typical of an early B star,
but are those of an accretion disk, which also would have an absolute
magnitude in the appropriate range.  In order to explain the
high reddening to the system, when $E(B-V)$ should be less than 0.4
for an object closer than 1kpc (Ishida 1969), CI98 invoke a circumstellar
origin.
Taking this into account, CI98 propose a new model for
the system where the primary is a compact object surrounded by
an accretion disk of $T_{\rm eff}\sim15000$K which hides the
central X-ray source.  The accretion disk mimics the atmosphere
of a B star, giving the observed spectrum.  

We have a number of objections against the distance and the model
which CI98 propose.  Firstly, we showed in section 2 that the spectral
type of the object is B0Ve.  For an accretion disk
to mimic such a spectrum it would have to have not only the temperature
($\sim 30000$K not $\sim15000$K) but the pressure 
($\sim 2\times10^3$ dyn cm$^{-2}$) and
structure of the outer layers of a B star, plus a region responsible
for producing Balmer series emission.  It is hard to imagine
such a disk forming, let alone being stable.  In addition we question why
such an object should have the typical colours of an $\sim 10000$ K 
accretion disk when it would have the spectrum of a much hotter disk.
Perhaps a better explanation for CI98 to have put forward would have been 
that
the system consists of a hot subdwarf plus compact object.  However,
as we have shown in section 2.1, the spectrum is definitely not that of
such an object either.

Next we consider our reddening measurements to the system.  As stated
previously, Howarth (1983) derived $E(B-V)\sim0.75$ from the 2200 {\AA}
feature in IUE spectra of the source.  We derive $E(B-V)\sim0.70$ from
our Na D$_2$ line measurements and 0.65 from the diffuse interstellar
band strengths.  The agreement between these three measurements
is striking.  It implies that the properties of the alleged circumstellar
material proposed by CI98 are identical to those of the interstellar
medium.  This strongly implies that the observed
reddening to LSI$+61^\circ$ 303 is interstellar, not circumstellar in origin.
We also note that 
Porceddu et al. (1992) have shown that excess
circumstellar extinction in Be stars does not produce strong diffuse 
interstellar bands.

Finally, CI98 propose the red $(B-V)_o$ and $(U-B)_o$ colours 
as an accretion disk
signature, although as we point out above it is that of a cooler
disk than they propose.  Comparison with the intrinsic colours of
normal B stars (Deutschmann et al. 1976) 
shows that they are $\sim 0.3$ magnitudes redder
in $B-V$ and $\sim 0.2$ magnitudes redder in $U-B$.
A more natural explanation of reddenings of this size 
is simply free-free emission from the circumstellar envelope
of the Be star (Schild 1983).
  
From all of the above we conclude that the simplest explanation is
that LSI+61$^\circ$ 303 contains a B0Ve star at a distance of
$\sim2$kpc plus a compact object as was previously thought,  
and that the {\it Hipparcos} distance is
somehow wrong.  We also note that the line of sight to 
LSI +61$^\circ$303 lies on the 
direction to the centre of the Cas OB6 association. 
Garmany \& Stencel (1992) give an
extinction corrected distance modulus of 11.9 
to that association, corresponding to a 
distance of 2.4 kpc. Therefore LSI +61$^\circ$303 is a likely member of 
Cas OB6.

\subsection{A0535+262}

We now consider the case of A0535+262.  The previous estimates of the
distance to this system were based on either the combination of a spectral
type and photometry (Hutchings et al. 1978, Giangrande et al. 1989,
Giovannelli and Graziati 1982) or Str\"{o}mgren photometry (Reig 
1996). They range from 1.3 to 2.4 kpc, the average being around
2kpc.  The reddening from the 2200 {\AA} feature is $E(B-V)\sim0.75$
(Giovannelli et al. 1980), and we showed in section 3.2 that a similar
reddening is derived from diffuse interstellar bands and the
Na D$_2$ line.

CI98 quote a {\it Hipparcos} distance to the system of 330 pc, 
with a 1$\sigma$ 
range of 210 to 780 pc.  They argue that the intrinsic $B-V$ of
the system is in the range $-0.13$ to $-0.30$, and therefore prefer a
spectral type of around B2.  They claim this is evidence against an
O9.7 spectral type, although such an object would have an intrinsic
colour only a few {\em hundredths} of a magnitude bluer.  
In addition they fail to take into account any reddening of
$(B-V)$ due to free-free emission (Schild 1983).  They also
state that the absolute visual magnitude corresponding to such
a distance is more consistent with an early B dwarf than a late
O giant.  Such a distance implies a low X-ray luminosity of
the system of only $\sim10^{33}$ erg s$^{-1}$.  CI98 propose that
systems with such low luminosities may in fact contain white dwarfs rather
than neutron stars, with perhaps only X Per standing out as
a neutron star system.

Once again we have several problems with these arguments.
Firstly the spectra presented in Fig. 3 are clearly those of
an O9.5-B0 star, and not a B2 object.  This shows the danger
of attempting to use colours to
derive spectral types of reddened objects (especially when they 
may show free-free emission!).  

Secondly we note that using a distance of 330pc and $E(B-V)=0.75$ we derive 
$M_V\sim-1$ for this object.  This is not consistent with an early
B dwarf as was stated by CI98, but with an object of spectral
type B6V (Deutschmann et al. 1976).  
It is also certainly not consistent with the
spectral type we derived in Section 2.

Next we again note the similarity of the three extinction measures presented
in section 3 as evidence that the reddening material is interstellar rather
than circumstellar.  Such large reddenings are incompatible with a distance
of only $\sim 330$pc (Ishida 1969).

Finally we point out that the explanation for the low X-ray luminosity
proposed by CI98 (white dwarfs rather than neutron stars in such systems)
does not hold for A0535+262.  X-ray spectra
of the source in outburst show cyclotron features
at around 50 and 100 keV (Kendziorra et al. 1994, Grove et al. 1995).  
These correspond to a magnetic field strength
of $\sim 10^{13}$ G (Araya \& Harding 1996), 
many orders of magnitude greater than
that possible for a white dwarf.  In addition both the measured spin-up of
the X-ray pulsations from the system (Li 1997) and
the presence of quasi-periodic oscillations (Finger et al. 1996) 
fit a neutron star model well.  The presence of a neutron star in this system
is therefore beyond any reasonable doubt.

From all of the above we conclude that the simplest explanation is
that A0535+262 contains a 09.7-B0 IIIe star plus a neutron star.
Using a spectral class of B0III ($M_V\sim-5.3$ - Vacca et al. 1996), 
a reddening $E(B-V)=0.75$,
and assuming a standard reddening law (Rieke \& Lebofsky 1985) 
we derive a distance of $\sim2$ kpc to the system.
We therefore again conclude that the `traditional' distance to the system
is more or less correct. 

\section{Conclusions}

We have shown that the {\it Hipparcos} derived distances ($\sim$ few hundred pc) 
to two Be/X-ray
binary systems, LSI +61$^\circ$ 303 and A0535+262 are inconsistent
with the spectral types, reddenings and apparent magnitudes of
the objects.  There is strong evidence that the `traditional' distances to
these objects (each $\sim 2$kpc) are in fact correct.
We note here that these two objects
have the worst goodness-of-fit values in the {\it Hipparcos}
catalogue (ESA 1997) of the CI98 sample (although they do lie below the maximum
``acceptable'' value of 3).  In addition they are the
faintest in the sample.   This appears to indicate that the 
application of the simple `goodness-of-fit' criterion that anything 
less than 3 is a good parallax to faint objects is not reliable, and 
that the interpretation of
{\it Hipparcos} parallax data should always be carried out with this in mind.
  
\section*{Acknowledgements}
Data reduction for this paper was carried out on the Southampton,
Liverpool John Moores and Sussex University
STARLINK nodes.  We thank Hannah Quaintrell for providing the INT spectrum 
of LSI+61$^\circ$ 303.  We thank Pablo Reig for his assistance in 
obtaining some of the JKT observations.
The JKT and INT are operated by
the ING on behalf of the UK Particle Physics and
Astronomy Research Council (PPARC) at the ORM Observatory, La Palma.   
The 1.5m telescope at Mount Palomar
is jointly owned by the California Institute of Technology and the Carnegie
Institute of Washington.  We thank Deepto Chakrabarty and Tom Prince for
their assistance in obtaining the Mount Palomar data.  
This research has
made use of the SIMBAD and Vizzier databases of the 
Observatorie Astronomique de Strasbourg and the La Palma data archive
of the Royal Greenwich Observatory.
Finally we thank Rob Fender for
bringing the Chevalier \& Ilovaisky paper to our attention.


\begin{thebibliography}{}
\bibitem[\protect\citename{null1}2000]{null1}
Allard F., Wesemael F., Fontaine G., et al., 1994, AJ 107, 1565
\bibitem[\protect\citename{ci3}1998]{ci3}
Araya R.A., Harding A.K., 1996, A\&AS, 120, C186
\bibitem[\protect\citename{ci3b}1998]{ci3b}
Baschek B., H\"{o}flich P., Scholz M., 1982, A\&A 112, 76
\bibitem[\protect\citename{ci4}1998]{ci4}
Bisscheroux B.C., Pols O.R., Kahabka P., et al., 1997, A\&A 317, 851
\bibitem[\protect\citename{ci5}1998]{ci5}
Chevalier C., Ilovaisky S., 1998, A\&A 330, 201 
\bibitem[\protect\citename{null6}2000]{null6}
Clark J.S., et al., 1998, MNRAS 294, 165 
\bibitem[\protect\citename{null7}2000]{null7}
Deutschmann et al., 1976, ApJS, 30, 97
\bibitem[\protect\citename{null7s}2000]{null7s}
Edwin R, 1988, RBS User Manual, La Palma User Manual 11, Royal
Greenwich Observatory
\bibitem[\protect\citename{null8a}2000]{null8a}
ESA 1997, The Hipparcos and Tycho Catalogues, ESA SP-1200 
\bibitem[\protect\citename{null8}2000]{null8}
Finger M.H., Wilson R.B., Harmon B.A., 1996, ApJ, 459, 288
\bibitem[\protect\citename{null8c}2000]{null8c}
Frail D.A., Hjellming R.M., 1991, AJ, 101, 2126
\bibitem[\protect\citename{null9}2000]{null9a}
Garmany C.D., Stencel R.E., 1992, A\&AS, 94, 211
\bibitem[\protect\citename{null9}2000]{null9}
Giangrande A., et al., 1980, A\&AS, 40, 289
\bibitem[\protect\citename{null10}2000]{null10}
Giovannelli F., Sabau Graziati L., 1992, Sp. Sci. Rev., 59, 1 
\bibitem[\protect\citename{null11}2000]{null11}
Giovannelli F., et al., 1980, in Proc. 2nd European IUE Conf., ESA SP-157, 159 
\bibitem[\protect\citename{null12}2000]{null12}
Granes P., Thom C., Vakili F., 1987, in Sletteback A., Snow T.P.,
eds, proc. 92nd IAU colloquium, Physics of Be Stars, Cambridge
University Press, p.66  
\bibitem[\protect\citename{null13}2000]{null13}
Gregory P.C., et al., 1979, AJ, 84, 1030
\bibitem[\protect\citename{null14}2000]{null14}
Grove, J.E., et al., 1995, ApJ, 438, 25
\bibitem[\protect\citename{null15}2000]{null15}
Herbig G.H., 1975, ApJ, 196, 129
\bibitem[\protect\citename{null16}2000]{null16}
Hobbs L.M., 1974, ApJ, 191, 381
\bibitem[\protect\citename{null17}2000]{null17}
Hoffleit D., Jaschek C., 1982, The Bright Star Catalogue, 4th
Edition, Yale University Observatory
\bibitem[\protect\citename{null18}2000]{null18}
Howarth I., 1983, MNRAS, 203, 801
\bibitem[\protect\citename{null19}2000]{null19}
Hunger K., Gruschinske J., Kudritzki R.P., Simon K.P., 1981, A\&A 95, 244
\bibitem[\protect\citename{null20}2000]{null20}
Hutchings J.B., et al., 1978, ApJ, 223, 530
\bibitem[\protect\citename{null21}2000]{null21}
Ishida K., 1969, MNRAS, 144, 55
\bibitem[\protect\citename{null23}2000]{null23}
Johnson H.L., 1966, ARA\&A, 4, 193
\bibitem[\protect\citename{null24}2000]{null24}
Kendziorra E., et al., 1994, A\&A, 291, L31
\bibitem[\protect\citename{null25}2000]{null25}
Li X., 1997, ApJ, 476, 278
\bibitem[\protect\citename{null25b}2000]{null25b}
McCarthy, J.K., 1988, PhD Thesis, California Inst. Tech.
\bibitem[\protect\citename{null26}2000]{null26}
Moehler S., Richtler T., de Boer K.S., et al., 1990, A\&AS 86, 53
\bibitem[\protect\citename{null27}2000]{null27a}
Paredes J.M., et al., 1994, A\&A, 288, 519
\bibitem[\protect\citename{null27}2000]{null27}
Popper D.M., 1980, ARA\&A, 18, 115
\bibitem[\protect\citename{null28}2000]{null28}
Porceddu I., Benvenuti P., Krelowski J., 1992, A\&A, 257, 745
\bibitem[\protect\citename{null29}2000]{null29}
Reig, P., 1996, PASP, 108, 639
\bibitem[\protect\citename{null30}2000]{null30}
Rieke G.H., Lebofsky M.J., 1985, ApJ, 288, 618
\bibitem[\protect\citename{null31}2000]{null31}
Robertson J.G., 1986, PASP, 98, 1220
\bibitem[\protect\citename{null32}2000]{null32}
Schild R., 1983, A\&A, 120, 223
\bibitem[\protect\citename{null33c}2000]{null33c}
Vacca W.D., Garmany C.D., Schull J.M., 1996, ApJ, 460, 914
\bibitem[\protect\citename{null34}2000]{null34}
Walborn N.R., Fitzpatrick E.L., 1990, PASP, 102, 379
\end{thebibliography}
\end{document}